\documentclass[longauth]{aa}  
\usepackage{graphicx}
\usepackage{amsmath}
\usepackage{amsfonts}
\usepackage{amssymb}
\usepackage{natbib}
\usepackage{txfonts}
\usepackage{tabularx}
\usepackage{subfig}

\begin{document}
\title{RoboTAP - target priorities for robotic microlensing observations}
\author{ M. Hundertmark\inst{\ref{R9}} \and R. A. Street\inst{\ref{R0}} \and Y. Tsapras\inst{\ref{R9}} \and E. Bachelet\inst{\ref{R0}} \and M. Dominik\inst{\ref{R3}}  \and K. Horne\inst{\ref{R3}} \and V. Bozza\inst{\ref{R10} \and \ref{R11}} \and D. M. Bramich \and A. Cassan\inst{\ref{R14}} \and G. D'Ago\inst{\ref{R23}} \and R. Figuera Jaimes\inst{\ref{R3} \and \ref{R2}} \and N. Kains\inst{\ref{R17} \and \ref{R2}} \and C. Ranc\inst{\ref{R22}} \and R. W. Schmidt\inst{\ref{R9}} \and C. Snodgrass\inst{\ref{R13}} \and J. Wambsganss\inst{\ref{R9}} \and I. A. Steele\inst{\ref{R4}} \and S. Mao\inst{\ref{R15}} \and K. Ment\inst{\ref{R24} \and \ref{R9}} \and J. Menzies\inst{\ref{R8}} \and Z. Li\inst{\ref{R0}} \and S. Cross\inst{\ref{R0}} \and D. Maoz\inst{\ref{R20}} \and Y. Shvartzvald\inst{\ref{R21} \and \ref{R20}}}
                                     
\institute{Las Cumbres Observatory Global Telescope Network, 6740 Cortona Drive, suite 102, Goleta, CA 93117, USA\label{R0} \and European Southern Observatory, Karl-Schwarzschild-Str. 2, 85748 Garching bei M\"unchen, Germany\label{R2} \and SUPA, School of Physics \& Astronomy, University of St Andrews, North Haugh, St Andrews KY16 9SS, UK\label{R3} \and Astrophysics Research Institute, Liverpool John Moores University, Liverpool CH41 1LD, UK\label{R4}  \and South African Astronomical Observatory, PO Box 9, Observatory 7935, South Africa\label{R8} \and Astronomisches Rechen-Institut, Zentrum f{\"u}r Astronomie der Universit{\"a}t Heidelberg (ZAH), 69120 Heidelberg, Germany\label{R9} \and Dipartimento di Fisica "E.R. Caianiello", Universit{\`a} di Salerno, Via Giovanni Paolo II 132, 84084-Fisciano (SA), Italy\label{R10}  \and Istituto Nazionale di Fisica Nucleare, Sezione di Napoli, Via Cintia, 80126 Napoli, Italy\label{R11} \and Planetary and Space Sciences, School of Physical Sciences, The Open University, Milton Keynes, MK7 6AA, UK\label{R13} \and Sorbonne Universit\'es, UPMC Univ Paris 6 et CNRS, UMR 7095, Institut d'Astrophysique de Paris, 98 bis bd Arago, 75014 Paris, France\label{R14} \and National Astronomical Observatories, Chinese Academy of Sciences, 100012 Beijing, China\label{R15} \and Space Telescope Institute, 3700 San Martin Drive, Baltimore, MD 21218, USA\label{R17} \and School of Physics and Astronomy, Tel Aviv University, Tel Aviv 69978, Israel\label{R20} \and Jet Propulsion Laboratory, M/S 169-506, 4800 Oak Grove Drive, Pasadena, CA 91109\label{R21} \and NASA Postdoctoral Program Fellow, NASA Goddard Space Flight Center, Mail Code 661, Greenbelt, MD 20771, USA\label{R22} \and INAF - Observatory of Capodimonte, Salita Moiariello, 16, 80131, Naples, Italy\label{R23} \and Department of Astronomy, Yale University, 52 Hillhouse Avenue, New Haven, CT 06511, USA\label{R24}}

   \authorrunning{M. Hundertmark et al.}
   \titlerunning{RoboTAP - target priorities for robotic microlensing observations}

   \date{}

  \abstract
   {The ability to automatically select scientifically-important
transient events from an alert stream of many such events, and to
conduct follow-up observations in response, will become increasingly
important in astronomy.  With wide-angle time domain surveys pushing
to fainter limiting magnitudes, the capability to follow-up on transient alerts 
far exceeds our follow-up telescope resources, and effective
target prioritization becomes essential.  The RoboNet-II microlensing
program is a pathfinder project which has developed an automated
target selection process (RoboTAP) for gravitational microlensing events which
are observed in real-time using the Las Cumbres Observatory telescope
network.}
   {Follow-up telescopes typically have a much smaller field-of-view compared to surveys, therefore the most promising microlensing events must be automatically selected at any given time from an annual sample exceeding 2000 events.
The main challenge is to select 
between events with a high planet detection sensitivity, aiming at the detection 
of many planets and characterizing planetary anomalies.}
   {Our target selection algorithm is a hybrid system based on 
estimates of the planet detection zones around a microlens. It 
follows automatic anomaly alerts and respects the expected survey 
coverage of specific events.}
   { We introduce the RoboTAP algorithm, whose purpose is to select and prioritize microlensing events with high sensitivity to planetary companions. In this work, we determine
the planet sensitivity of the RoboNet follow-up program and provide a working
example of how a broker can be designed for a real-life transient science program 
conducting follow-up observations in response to alerts, exploring the issues that
will confront similar programs being developed for the Large Synoptic Survey Telescope (LSST) and other time domain surveys.}
   {}

   \maketitle
%
   
\section{Introduction}

\subsection{RoboTAP in context}

In an era of increasing sky coverage and telescope \'{e}tendue, surveys monitor billions of stars and alert whenever an astronomical object changes its brightness. Most science cases will require follow-up observations responding to these alerts. This is especially true for transient events that, by definition, occur only once. The sheer number of alerts issued by the Large Synoptic Survey Telescope (LSST), for example, can be on the order of one million per night \citep{rid14} and therefore can not be handled by humans. Brokers for processing and filtering the alert stream for rapid response telescopes already exist, for instance the Arizona-NOAO Temporal Analysis and Response to Events System (ANTARES) \citep{sah16}, but more work is needed to address the variety of science cases. The demand for such systems will increase due to emerging projects such as the Evryscope \citep{law15}, the Zwicky Transient Facility \citep{smi14,bel14}, the BlackGEM survey and last but not least the Large Synoptic Survey Telescope \citep{lss09}. The RoboTAP system is such an automated broker system and a prototypical example for a range of transient science cases. 

The philosophy behind the RoboTAP implementation is to provide a robust and computationally lightweight algorithm that only takes seconds for an event feed with $\sim1000$ events. In order to achieve that, we followed a small set of basic working principles. We start by carrying out computationally inexpensive operations first and apply increasingly complex criteria later; we replace estimates that are only required to be roughly known by suitable interpolants. Incoming event parameters are kept and updated in memory all the time. The same technical considerations apply to selecting a suitable and algebraically simple priority function.

\begin{figure}
\resizebox{\hsize}{!}{\includegraphics{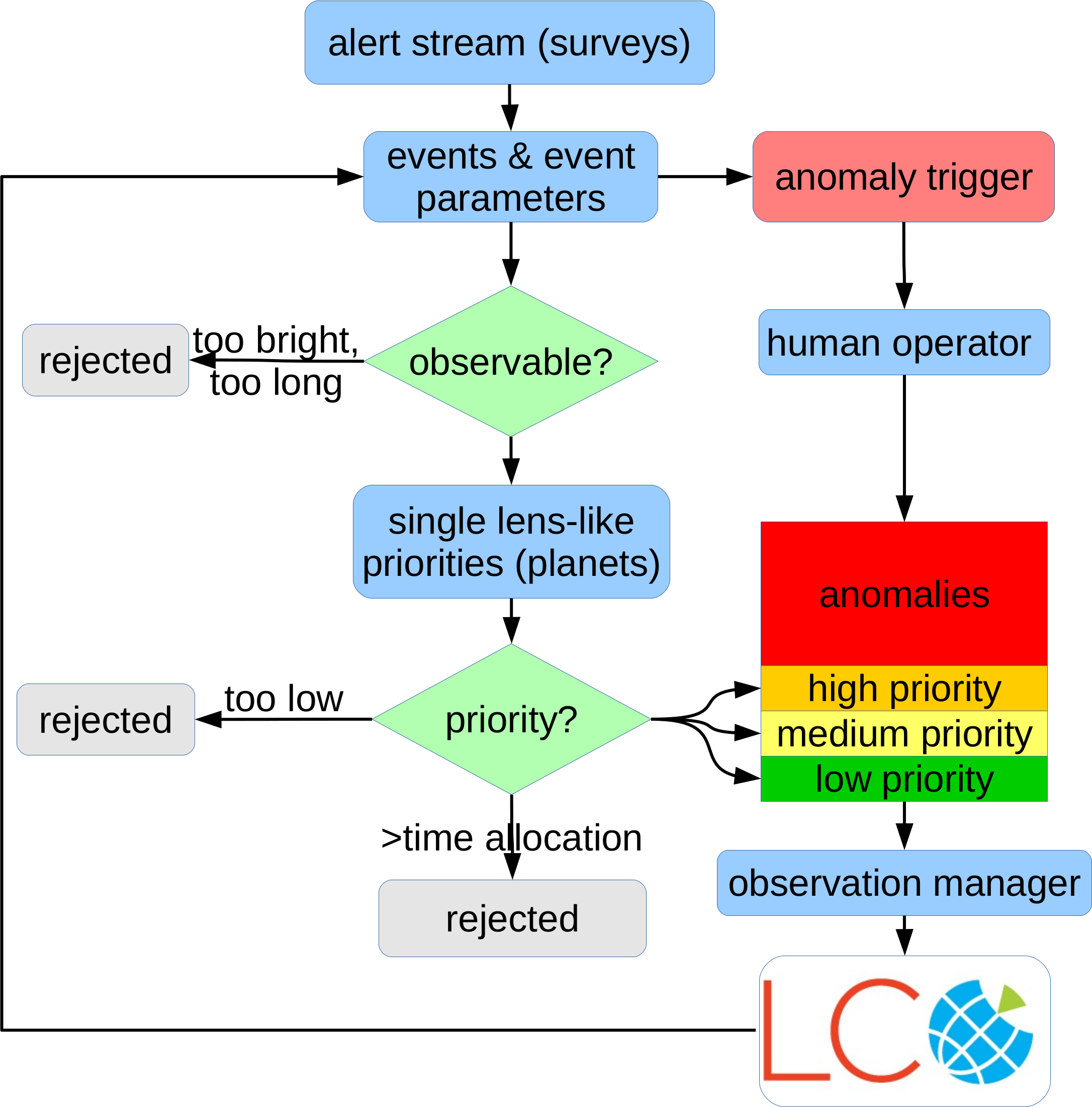}}
\caption{Workflow of the RoboTAP system processing the incoming alert stream of regular and anomalous microlensing events from ARTEMiS. \label{fig:flowchart}}
\end{figure}

Fig.~\ref{fig:flowchart} depicts the simplified workflow of RoboTAP acting as a broker between the event stream and the observation manager. The main source for the alert stream are existing microlensing surveys such as the Optical Gravitational Lensing Experiment (OGLE) survey \citep{uda94} or the Microlensing Observations in Astrophysics (MOA) survey \citep{bon04}. The workflow illustrates how an incoming stream of events is subjected to filters and triggers observations which in turn can refine the event parameters in a closed-loop system. It also highlights that further building blocks are necessary to efficiently process the event stream. Only events that have been classified as microlensing are processed. Moreover, we introduce two sub-classes - {\it regular} and {\it anomalous} microlensing events. For that purpose, we rely on the ARTEMiS system \citep{dom08} providing us with event parameters and anomaly triggers for events with irregular light curve shapes. In this two-tier approach, we generate preliminary target lists by reducing the number of transient (microlensing) events based on observability, survey coverage and expected duration. Finally, we rank all events based on our priority function. Anomalous and non-anomalous are allocated a pre-defined fraction of the observing time on the network.The corresponding target lists are then formatted as separate observing requests and submitted to the telescope network.

\subsection{Related observatories}

\begin{figure*}
\resizebox{\hsize}{!}{\includegraphics{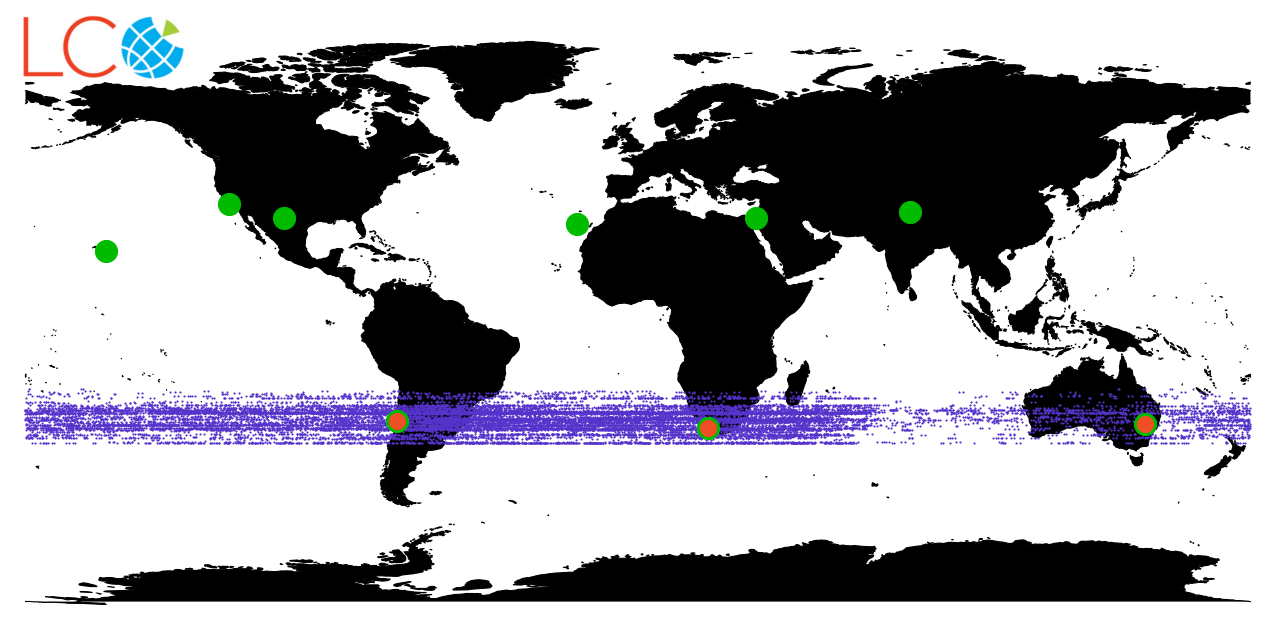}}
\caption{Sites hosting LCO telescopes (green) and LCO sites used for the Key project (red) are shown along with observed microlensing target positions at zenith over Earth for the 2015 season (blue dots). The Australian site hosts only two telescopes which leads to less coverage, otherwise the event distribution would be homogeneous along all longitudes.\label{fig:lco}}
\end{figure*}

While our approach is of general interest for all astronomical observatories, the actual implementation was customized to the Las Cumbres Observatory. The Las Cumbres Observatory (LCO) is an organization carrying out astronomical research and operating a global network consisting of homogeneous telescopes of different aperture classes \citep{bro13}. In addition to the two 2m Faulkes telescopes originally used by the RoboNet collaboration \citep{bur07,sno08,bra08,tsa09,str11}, $9\times1\textrm{m}$ telescopes became available in 2013 to the microlensing program of which 8 are on sites in the Southern Hemisphere, where most of the microlensing targets lie. The geographical location of telescopes as shown in Fig.~\ref{fig:lco}, ensures that targets can be observed nearly seamlessly if weather permits. The lower density of observations above Australia is a consequence of having only $2\times1\textrm{m}$ telescopes available as well as the characteristic weather pattern.

With an emphasis on robotic time-domain astronomy, gravitational microlensing perfectly fits into the observing program of LCO. Moreover, with its telescope clusters in Chile, South Africa and Australia, it achieves unprecedented coverage of events in the Galactic center. Consequently, microlensing was selected to be part of the 2013 pilot program of the 1m network and was awarded Key Project\footnote{PI: R.A.~Street} status on the LCO network between 2014--2016. At the beginning of the 2013 pilot phase the multi-purpose scheduler was not available and the telescopes were initially equipped with SBIG STZ-16803 cameras\footnote{\url{http://LCO.net/network/instrumentation/1m-sbig-camera}}. The Key Project was interrupted by the 2015 Spitzer campaign as described by \cite{yee15} for obtaining simultaneous ground and space-based observations of microlensing events. This paper will focus on the pilot phase and the first year of operations with its lower number of competing projects, direct scheduling and corresponding camera systems. 

\subsection{Gravitational Microlensing}

Gravitational microlensing is one of the well established methods of exoplanet detection, probing the Galactic planet population down to Earth-mass range with ground-based observations \citep{ben96, wam97}. Light from a distant source star is gravitationally deflected by a foreground lens star, producing an increase of brightness of the source from the point of view of the observer. Planets orbiting faint (even unseen) lens stars can be discovered in this manner, because the light curve shape is related to the mass of the lens-system and depends on the mass ratio between planet and host star as well as their angular separation. Moreover, the planet population beyond the snow-line can be probed due to the increased sensitivity of this detection technique \citep{par06}. This is the main science driver of the aforementioned Key Project.

The sample of microlensing events and their absolute sensitivity to planets provides a statistical and independent measure of the planet abundance in our Galaxy \citep{sno04,gau02,gou10,sum10,sum11,cas12}. The detection of microlensing events itself is subject to statistics, as only $\sim1$ in a million stars in the Galactic bulge is sufficiently aligned with a deflecting foreground star as to ensure that more light will be redirected to the observer as predicted by the seminal paper of \cite{pac86}. Both planet detections and non-detections provide important constraints on the frequency and distribution of planets in the Galaxy.

Most microlensing events follow a characteristic symmetric light curve which is a direct consequence of Einstein's deflection angle. This deflection can be described by a simple lens equation, relating source position, image position and deflection angle. The solutions to the lens equation provide the image positions. The radius of the ring-like image in a co-linear lens-source-observer configuration defines the Einstein radius

\begin{equation}
\theta_{\mathrm{E}} = \sqrt{\frac{4 G M_{\mathrm{L}}}{c^2} \left(D^{-1}_{\mathrm{L}}-D^{-1}_{\mathrm{S}}\right)}
\end{equation} 
constraining the typical angular scale of the effect, where $D_{\mathrm{L}}$ denotes the distance from observer to deflecting lens of mass $M_{\mathrm{L}}$ and $D_{\mathrm{S}}$ denotes the distance to the source star.
The total observable brightness increase is time dependent due to the relative source-lens proper motion. While multiple images can be detected and resolved for massive, extra-galactic source-lens configurations, \textit{micro}lensing on Galactic scales means that the image separation is of the order of  $\sim1\,\mathrm{mas}$ and remains unresolved in the optical regime with current imaging technology. The total flux $F(t)$ of a microlensing event with a contribution of source flux $F_{\mathrm{S}}$ and blend flux $F_{\mathrm{B}}$ can be written as
\begin{equation}
F(t) = F_{S} A \left(u \left(t\right)\right) + F_{B},
\label{eq:fluxeq}
\end{equation}
where the magnification $A$ of a single lens is obtained by adding the differential magnifications of each image $A_{\pm}$ which depends on the actual image positions in units of the angular Einstein radius
\begin{equation}
u_{\pm}=\frac{u {\pm} \left(u^2+4\right)^{1/2}}{2}.
\end{equation}
The image magnification for the minor image $A_{-}$ and the major image $A_{+}$ is
\begin{equation}
A_{\pm}=\frac{u_{\pm}^2}{u_{+}^2-u_{-}^2}.
\end{equation}
For practical purposes, we use the blend ratio 
\begin{equation}
g= \frac{F_{B}}{F_{S}},
\end{equation}
and express the lens position in units of the angular Einstein radius. The time-dependent source-lens distance denoted as $u=u(t)$ defines the shape of the point-source point-lens (PSPL) Paczy{\'n}ski light curve 
\begin{equation}
A\left(u\right) = A_{+}+A_{-}=\frac{u^2+2}{u\sqrt{u^2+4}}
\label{eq:pspl}
\end{equation}
in units of the angular Einstein radius. The latter expression can be derived analytically, because the lens equation of a single object can be solved for the image positions. 
The relative lens-source proper motion plays a role for the light curve shape but for events that do not exhibit parallax effect \citep{gou92} a uniform motion along a straight line serves as reasonable approximation on time scales of several days or weeks with the source-lens distance
\begin{equation}
u=\sqrt{u_0^2 + \left(\frac{t-t_0}{t_{\mathrm{E}}}\right)^2}.
\label{eq:track}
\end{equation}
The time of maximum magnification is denoted as $t_0$, the Einstein-radius crossing time is $t_{\mathrm{E}}$ and the minimum separation of lens and source is $u_0$. As an analytic model for all further considerations, Eq.~\ref{eq:pspl} plays a crucial role for prioritizing events and detecting anomalies. In this context, we consider each event as anomalous if it deviates from the standard point-source point-lens (PSPL) model as defined in Eq.~\ref{eq:pspl} and is triggered by the anomaly detector algorithm\footnote{in a typical scenario more than 5 clearly deviating points from at least 2 consecutive nights are required to confirm an anomaly.} described by \cite{dom07}. 

The science driver of the LCO Key project ``Exploring Cool Planets Beyond the Snowline'' is to find more planets with orbits between 0.5 and 10\,AU which corresponds to the orbital region where the temperature drops below the freezing point of water. Gravitational microlensing provides us with mass measurements and the projected orbital radii, thus populating the mass vs. orbital radius diagram. Microlensing is complementary to the transit and the radial-velocity methods, since it is sensitive to Earth mass planets beyond the water snow line.

\section{Prioritizing microlensing events}

\subsection{Priority and planet detection zones}

Microlensing follow-up programs can be adapted to different scientific objectives such as studying the brown-dwarf population \citep{str13}, providing mass measurements of distant stellar remnants \citep{mao02,wyr16} or searching for planets beyond the habitable zone \citep{tsa14}. Searching for planets beyond the snowline is the main motivation behind the RoboNet-II program for following-up microlensing events with telescopes from the Las Cumbres Observatory Global Telescope Network (LCO) and the Liverpool Telescope \citep{bro13,ste04}. 

\begin{figure}
\resizebox{\hsize}{!}{\includegraphics{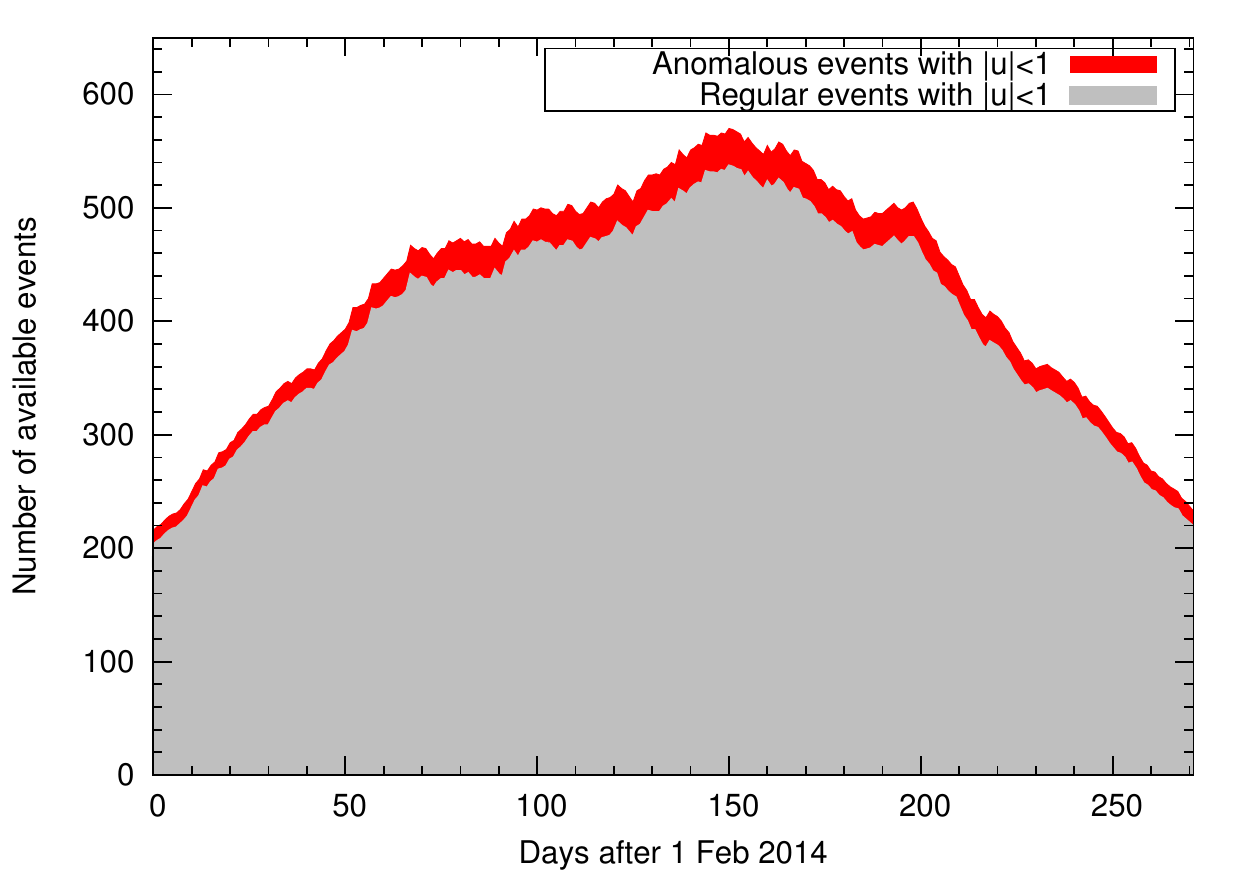}}
\caption{Number of ongoing regular and anomalous events in the season 2014 before and after $t_0$.\label{fig:ongoingevents}}
\end{figure}

For the pilot phase of the 1m network in 2013 we combined the approach of prioritizing events based on planet probability estimates with network specific observational constraints \citep{hor09,dom10}. This approach addresses the challenge of selecting the right targets out of hundreds of ongoing events. For reasons of simplicity, one often considers an event to be ongoing when the separation between lens and source is smaller than the Einstein radius and the source thus experiences a brightening by at least 0.3\,mag with respect to baseline. In practice, we permit observations for source-lens distances $\left|u\right|<1.5$ when an unblended source appears to be 0.1\,mag brighter. Fig.~\ref{fig:ongoingevents} illustrates the number of available events at any time during the microlensing season (April to October). These 200-600 events include 5-10\% anomalous events deviating from a simple static point lens and are reported by anomaly detectors. Examples of different anomaly types can be found in \cite{tsa16}. Not all anomalous events can be monitored continuously, and even with two-to-three follow-up telescopes per site we can only cover a handful of events with high enough cadence. A suitable prioritization is therefore required. 

\begin{figure*}
\resizebox{\hsize}{!}{\includegraphics{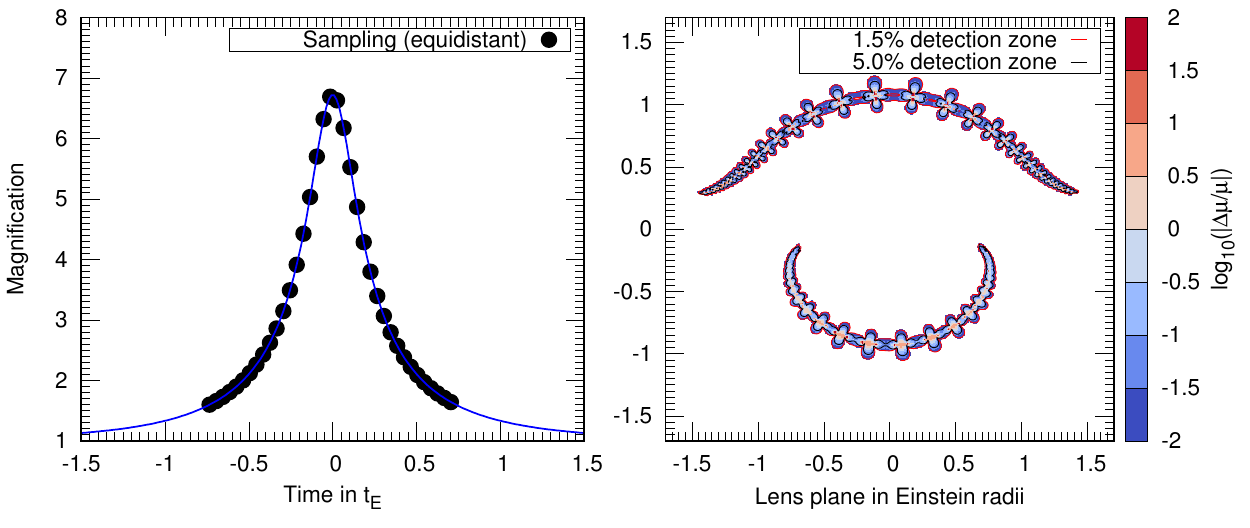}}
\caption{Left: the light curve of a standard point lens light curve with a given equidistant sampling rate of 1 visit per night is shown. Right: the corresponding planet detection sensitivity of planets perturbing the light curve is shown for the lens plane assuming a low-mass planet with mass ratio $q=10^{-4}$. \label{fig:dza}}
\end{figure*}

The idea of defining the planet detection probability of a given microlensing event via its detection zone goes back to the initial suggestions how likely it is that planets orbiting microlenses can be found \citep{gou92b}. A single-lens microlensing event creates a pair of images propagating through different parts of the lens plane where they can be perturbed by the presence of a planet. Estimating when such a perturbation becomes photometrically detectable gives the chance of finding such a planet. Fig.~\ref{fig:dza} illustrates the locations, where the extra planetary deflection is powerful enough to generate detectable deviations. The detection requires comparison with an underlying single-lens model, but the detection zone can shrink or grow depending on photometric accuracy. Therefore, one more commonly uses a $\Delta \chi^2$ threshold. For reporting a discovery one would require that observations lead to a change in $\Delta \chi^2>25$, corresponding to $5\,\sigma$ detections, if the underlying uncertainties are appropriately assigned and follow a Gaussian distribution. The complicated way how uncertainties are affected by crowded field photometry \citep{bac15} frequently leads to a distribution of residuals which differs from that assumption and needs rescaling in the form of 
\begin{equation}
\sigma_{\mathrm{rescaled}}^2= k \left(\sigma_{0}^{2} + \sigma_{\mathrm{reported}}^2\right),
\label{eq:rescale}
\end{equation}
where $\sigma_{\mathrm{reported}}$ refers to the original uncertainties and the parameters $k,\sigma_0$ are chosen so that the reduced $\chi^2$ becomes one for each independent light curve. Since some light curves have uncertainties that can be off by a factor of 2, one commonly requires a threshold of $\Delta \chi^2>100$ for a detection. 

In order to quickly calculate the planet detection probability $\Psi$, the exposure time $t_{\mathrm{exp}}$ is adjusted according to the expected brightness. We thereby ensure that a photometric accuracy of 1.5 to 5\% is reached and a planetary perturbation is unlikely to be missed. One of the main changes between the empirical relations by \cite{hor09} and \cite{dom10}, which are both based on the planet detection zone area, is the underlying logarithmic distribution of planets along the orbital axis which is also consistent with observations \citep{cas12}. The planet probability $\Psi$ for major and minor image positions $u_{\pm}$ and their corresponding magnification $A_{\pm}$ is given by:
\begin{equation}
  \Psi\left(u\right)=\frac{2 A_{+} -1}{u_{+}^{\gamma}}+\frac{2 A_{-}}{u_{-}^{\gamma}},
 \end{equation}
where the parameter $\gamma$ is chosen to be $2$ for the assumed population of planets, which comes from the aforementioned empirical relations combined with the assumption of a logarithmically distributed orbital axis. For $\gamma=2$ the detection probability can be simplified:
\begin{equation}
\Psi\left(u\right)=\frac{4}{u \sqrt{u^2+4}} -\frac{2}{u^2+2+u\sqrt{u^2+4}}.
\label{eq:psi}
\end{equation}
We consider the gain of observing an event to be given by the ratio of the planet detection probability and the invested observing time for a photon-noise dominated light curve. The priority function then reads
\begin{equation}
\Omega_{S} \propto \frac{A^{3/2} \Psi}{\left(A+F_{B}/F_{S}\right)\left(1+F_{B}/F_{S}\right)}.
\label{eq:omega}
\end{equation}
The latter expression makes implicitly the assumption that the suggested exposure time can be reached, the noise model is correct and that the underlying magnification $A$ is sufficiently constrained by fitting a point-source point-lens model to the data. In the next section we will discuss to what extent these assumptions affect our results. One of the strategic differences between RoboNet and the strategy implemented in ARTEMiS is the absence of spending observing time for checking events when they slightly deviate from a point-source point-lens model. Based on estimates of earlier seasons we expect that the priority function alone will deliver and select high priority events that eventually will turn out to be anomalous - including those of potential planetary nature. Because of the two-step multi-tier strategy, it is more difficult to evaluate the sensitivity of our follow-up program. In order to ensure that anomalous events are not completely dominating our Key program, we have put a cap on the fraction of allocated anomalies.

To get a better understanding of where microlensing can find planets, we illustrate the detection zone in the right panel of Fig.~\ref{fig:dza}. For each data-point a single point lens creates two images sweeping along the Einstein ring. Depending on the proximity of these images to a potential planet the light curve can be perturbed. For planets at various positions the corresponding deviation can be recorded and if it exceeds a certain threshold a planetary anomaly can be detected. The threshold depends on the achievable photometric accuracy as well as the desired number of deviating points. We assume that planets deviating by more than 1.5\% (0.015 mag) can be detected if the exposure time can be adjusted accordingly.

For typical host stars ranging from 0.1 to 0.5\,$M_{\odot}$ we simulate the detection zone area for a mass ratio of $q\approx 10^{-4}$. This corresponds to planets in the range of 3 to 17 Earth masses covering super-Earths and mini-Neptunes. For typical point lens parameters ($u_0=0.15, t_{\mathrm{E}}=25\,\mathrm{d}, \rho=10^{-3}\,\theta_{\mathrm{E}}$) the 1.5\% detection zone area shown in Fig.~\ref{fig:dza} also accounts for overlapping detection zones covering the same region. Finite source effects have been considered in the form of a uniform disk using the binary lens model by \cite{boz10}\footnote{\url{http://www.fisica.unisa.it/gravitationAstrophysics/VBBinaryLensing.htm}}. Most light curves can be simulated under the assumption that the source star is an infinitesimally small box and the distortion of that box is given by the inverse determinant of the lens equation. Whenever an image approaches a so-called critical line, the change of magnification is so dramatic that the surface of the source star is differentially magnified, which also ensures that the predicted brightness change remains finite.

\cite{dom09} hints at the unpredictability of events or in other words the difficulty of predicting the light curve days before the event reaches its maximum magnification. To successfully prioritize events, we expect that order relations 
\begin{equation}
\Omega_{\mathrm{S},i} > \Omega_{\mathrm{S},j}
\end{equation}
hold for each pair of consecutive events $i,j$. A heuristic estimate can be achieved assuming that the uncertainty of priority is limited by the uncertainty of $u$ as Eq.~\ref{eq:omega} monotonically increases with $u$. If the peak has not been reached and the blend flux is low, it is safe to assume that $u_0\in\left[0,F_{\mathrm{now}}/F_{\mathrm{S}}\right]$. The corresponding uniform distribution and its variance $\Delta u/12$ can be used to estimate the uncertainty in $u$ and for the last reported flux $F_{\mathrm{now}}$ that provides an upper bound for $u$ that gives the uncertainty
\begin{equation}
\sigma_{u}^2 \approx \frac{F^2_{\mathrm{now}}}{12 F^2_{\mathrm{S}}}. 
\label{eq:uncpre}
\\\end{equation}

If multiple events are considered before reaching their peak, those with the highest magnification estimate have automatically the lowest uncertainty and as long as the available events are separated by a factor of $1/\sqrt{12}$ in $u_0$ the priority order is appropriate and that corresponds to the aforementioned variance of a uniform distribution. For events with $t>t_0$, $u$ is much better constrained. After observing the peak on a single occasion with a relative accuracy of $\delta_F=1.5\,\%$ and by estimating that $A \approx 1/u$ and $A_{\mathrm{true}}>A_{\mathrm{peak}}$ 
\begin{equation}
\sigma_{u}^2 < \delta_F^2 u^2.
\end{equation}
This can be interpreted by comparison with Eq.~\ref{eq:uncpre} for $F_{\mathrm{now}} \approx F_{\mathrm{peak}}$ where the uncertainty $\sigma_u$ roughly follows $1/u$ before the peak and $u$ after the peak. 

Before the peak is characterized, it is difficult to optimize the exposure time regardless of our knowledge of $g$. After the peak, a higher underlying signal-to-noise ratio in flux units is required where events with $g \gg 10$ may end up with more than 1\,\% noise in $A$. 

\subsection{Simulating a fiducial season} 

\begin{figure}
\resizebox{\hsize}{!}{\includegraphics{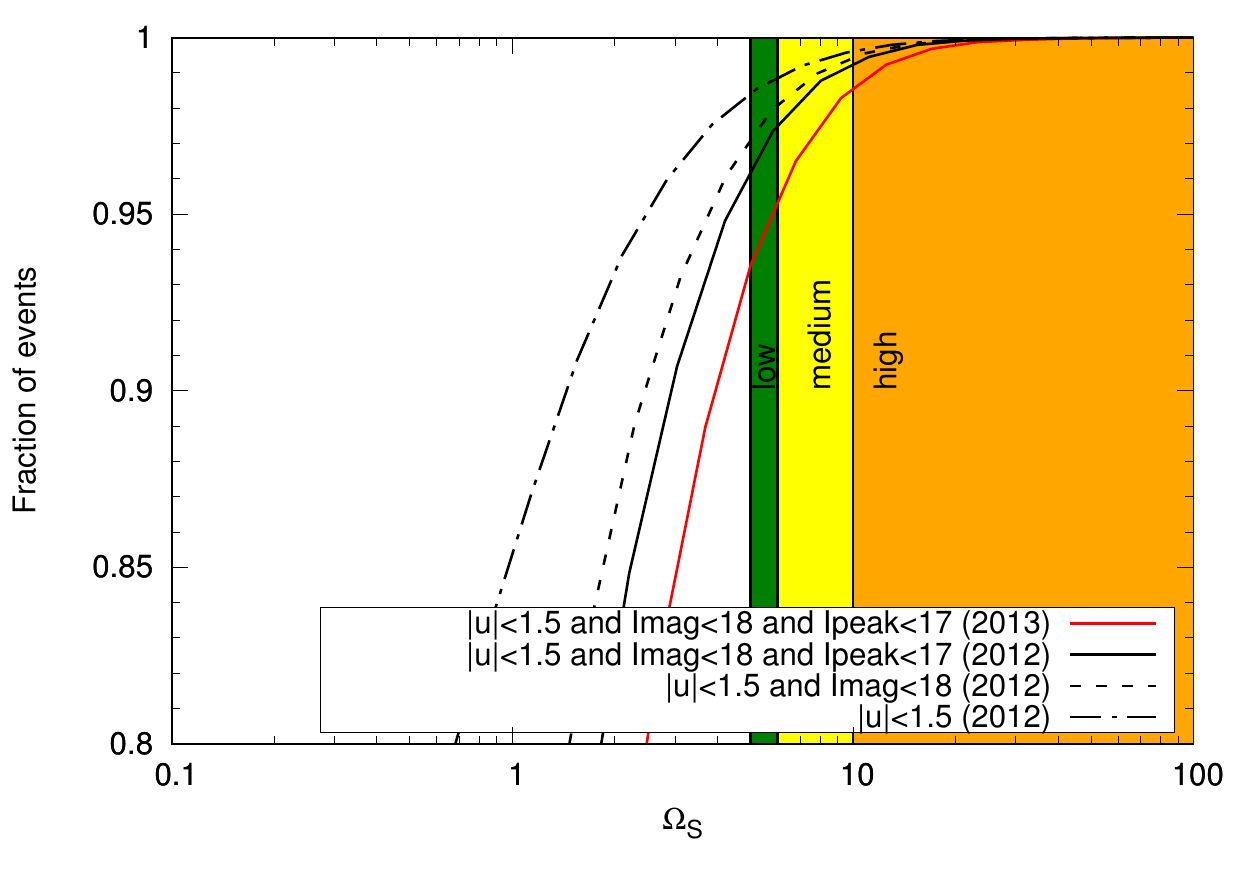}}
\caption{Cumulative priorities based on simulated events matching the season 2012 are shown for the underlying observing constraints and after the season 2013 has finished. \label{fig:priorities}}
\end{figure}

Before analyzing the outcome of real observations we shall provide an estimate of the expected number of events and achievable number of planets based on a simple model of basic microlensing parameters. For that purpose, we simulate ensembles of events with event parameters $t_{\mathrm{E}},u_0,I_{\mathrm{baseline}},g$ where samples are drawn using a Gaussian kernel density estimate of PSPL parameters obtained from post-season ARTEMiS fits of the 2012 season. We will show at a later point that the real-time ranking is sufficiently close to the post-season estimate. For each event we simulate a random time in the season and let the system determine its current priority state. Based on that simulation we determine the priority thresholds for the season. The cumulative distribution of the priority function as shown in Fig.~\ref{fig:priorities} indicates that 1\%,2\% and 1\% of events would fall in our {\it low}, {\it medium} and {\it high} priority categories. Consequently, we would expect to request observations of 20, 40 and 20 non-anomalous events assuming that 2000 events are provided by survey teams. 

In practice, the assumed number of underlying events turned out to be too low, because the revised simulation based on 2013 season parameters hints at an excess of low-priority targets. One explanation for the corresponding offset is a new discovery channel of the Optical Gravitational Lensing Experiment (OGLE-IV) in its fourth phase \citep{uda15} reporting fainter microlensing events which has lead to an increase of 15\% in terms of discovered events, so effectively we should expect to see 40,70 and 30 events within our thresholds. Fainter source stars that are magnified beyond the limiting magnitude are by definition of higher magnification and thus more likely to contribute to events with higher priorities. 

Based on the confirmed planets on the exoplanet archive\footnote{\url{http://exoplanetarchive.ipac.caltech.edu}}, we have seen 6 planetary events with 7 planets in the OGLE fields for 2012 and about 1700 stellar microlensing events in total. One event contains two planets \citep{han13}. With no assumption on the optimal choice for selecting events, we have a 1/5 chance to see a single planet when covering 50 events. Our choice of parameters and detection probability $\psi$ can be calculated at the peak of each light curve which is an approximation of the planet detection sensitivity of each event. That gives us a relative number of how many more planets we expect to see. 

Our strategy automatically incorporates that approach by assigning a priority level $\Omega_S$. The planets shown are based on the parameter records and not on published values to give us an indication whether the system would have picked the target. While 99.9\% of the peak planet detection probability is concentrated in 100 targets, two thirds of published planets instead belong to the 200 highest priority targets. We have limited that estimate to data from 2013 onwards, because older data would misrepresent the input from the anomaly detection and available survey data. Planets in the wings can still be detected although the planet detection probability is several orders of magnitude lower, if it is sampled in a way that prevents overlapping detection zone areas for many events.

\subsection{Characterizing events}
\begin{figure*}
\resizebox{\hsize}{!}{\includegraphics{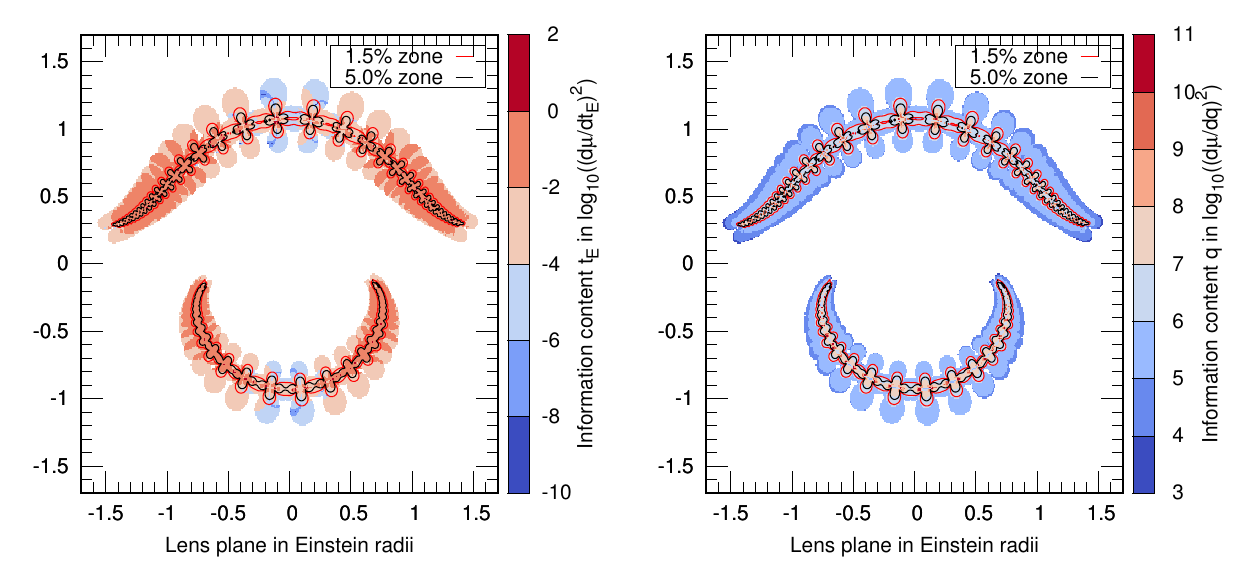}}
  \caption{The information content for characterizing $t_{\mathrm{E}}$ (left) and $q$ (right) is shown for an unblended $q=10^{-4}$ light curve with known parameters. Each of the color-coded points corresponds to a potential planet and its maximum contribution for the given sampling (cf. Fig.~\ref{fig:dza}). The mass ratio is optimally characterized by exactly following the detection zone area, while the Einstein time $t_{\mathrm{E}}$ could benefit from more observations in the wings. For 1\,\% photometric accuracy, the color-coded logarithmic sensitivity just needs to exceed 1 in order to ensure 10\% accuracy in each parameter. Since information is additive, one can estimate the number of required points.}
  \label{fig:bininfo}
\end{figure*}
The most important physical parameter that is accessible through gravitational microlensing is the planet mass:
\begin{equation}
M_{\mathrm{p}}=\frac{\mu t_{\mathrm{E}}}{\kappa \pi_{\mathrm{E}}}\frac{q}{1+q}; \kappa \approx 8.14\frac{\mathrm{mas}}{M_{\odot}},
\end{equation}
where $\mu$ denotes the lens-source relative proper motion, $\pi_{\mathrm{E}}$ is the magnitude of the microlensing parallax vector expressed in units of the Einstein radius, $q=M_{\mathrm{p}}/M_{\mathrm{host}}$ is the mass ratio and $\kappa$ a constant \citep{gou00}. Propagating the uncertainties and writing all uncertainty contributions relative to their quantities yields:
\begin{equation}
\sigma_{M_{\mathrm{p}}} \approx \frac{\mu q t_{\mathrm{E}}}{\kappa \pi_{\mathrm{E}} \left(1+q\right)} \left( \frac{\sigma_{t_{\mathrm{E}}}^2}{t_{\mathrm{E}}^2}+\frac{\sigma_{\pi_{\mathrm{E}}}^2}{\pi_{\mathrm{E}}^2}+\frac{\sigma_{\mu}^2}{\mu^2} + \frac{\sigma_q^2}{q^2\left(1+q\right)}\right)^{1/2}.
\end{equation}
For small mass ratios $1/(1+q)$ is sufficiently close to one and thus the planet mass uncertainty is dominated by the term with the largest relative uncertainty. In the following, we will focus on the uncertainties of mass-ratio and timescale since these can be constrained better for most events. In order to quantify how the uncertainties in the fit-parameters can be reduced, we use the total information content expressed in the Fisher matrix:
\begin{equation}
I_{i,j}(t)= \left<\left(\frac{\partial \log \mathcal{L}}{\partial p_i} \right)\left(\frac{\partial \log \mathcal{L}}{\partial p_j} \right)\right>,
\label{eq:fismat}
\end{equation}
which is the inverse of the covariance matrix depending on likelihood $\mathcal{L}$ and estimated parameters $p_i$. The expected value of this expression can be used to calculate how each data point contributes to the event parameters extracted from a fiducial model fitted to the light curve. The Fisher matrix simplifies in the $\chi^2$-case for a given time $t$ to:
\begin{equation}
I_{i,j}(t)= \frac{1}{\sigma^2_f}\left(\frac{\partial f_{\mathrm{model}}}{\partial p_i}\right)\left(\frac{\partial f_{\mathrm{model}}}{\partial p_j}\right)
\label{eq:fismat2}
\end{equation}
Evaluating this expression requires only the best-fit model $f_{\mathrm{model}}$ and the derivatives with respect to the parameters $p_i$. The information content of the full light curve is the sum of all matrices $I_{i,j}$. The microlensing model is usually given in the form shown in Eq.~\ref{eq:fluxeq} and thus depends on $F\left(A\left(u\left(p\right)\right)\right)$. \cite{hor09} and \cite{dom10} focus on finding planets rather than characterizing parameters. Fig.~\ref{fig:bininfo} shows areas associated with high probabilities of planet detection, for a typical detection threshold of 1.5\% in photometric accuracy and mass ratios $q>10^{-4}$. 

The sensitivity levels shown in Fig.~\ref{fig:bininfo} can be understood as inverse parameter uncertainties. In case of the mass ratio $q$, it follows directly the detection zone area, while for the Einstein time one needs a better coverage in the wings. The exact location of information maxima is not known before the peak and based on the distribution of these maxima a sampling interval of at least $t_{\mathrm{E}}/10$ is advisable. That holds for achieving about 10\,\% accuracy in $q$, but there is an obvious caveat. The corresponding measurements need to be affected by the planet. For that purpose, our highest sampling interval is 15\,min and follows $A^{1/2}t_{\mathrm{E}}$ \citep{hor09,dom10}. We would also like to highlight, that the azimuthal lobes in Fig.~\ref{fig:dza} overlap for the equidistant sampling. The radial detection zone lobes do not overlap to the same extent, which explains why the increased sampling rate at the peak is beneficial.

For low-mass planets we expect only a small perturbation in the light curve. Therefore, we estimate the contribution to the timescale based on a PSPL model. If the sampling rate would aim to maximize our understanding of $t_{\mathrm{E}}$ the sampling strategy needs to be adjusted. The sensitivity with respect to a parameter $p$, such as the Einstein time $t_{\mathrm{E}}$, can be obtained from
\begin{equation}
\frac{\partial F_{\mathrm{PSPL}}}{\partial t_{\mathrm{E}}} = F_{\mathrm{S}} \frac{\partial A}{\partial u}\frac{\partial u}{\partial t_{\mathrm{E}}},
\end{equation}
where
\begin{equation}
\frac{\partial A}{\partial u} = \frac{-8}{\sqrt{u^2+4}\left(u^4+4u^2\right)},
\end{equation}
and the inner derivative is
\begin{eqnarray}
\frac{\partial u}{\partial t_{\mathrm{E}}} &=&\frac{-(t-t_0)^2}{t_{\mathrm{E}}^3 u}.
\end{eqnarray}
This implies that the sensitivity vanishes at $t=t_0$. The magnification $A$ can be approximated as $A \approx 1+ \left(3 u^2 \right)^{-1}$ \citep{hor09} which leads to a simple estimate of  sensitivity:
\begin{equation}
\frac{\partial F_{\mathrm{PSPL}}}{\partial t_{\mathrm{E}}} = 2 F_{\mathrm{S}} \frac{\left(t-t_0\right)^2}{3 t_{\mathrm{E}}^3 u^4}
\end{equation}
The latter expression has two implications for the sampling rate: to characterize $t_{\mathrm{E}}$ it should follow roughly $\left(t^3_{\mathrm{E}} u^4\right)^{-1}$ and the sampling interval $T$ should be scaled with $t_{\mathrm{E}}$. For the 2013 observing season, the requested sampling rate was kept independent of the Einstein time, but in 2014 $t_{\mathrm{E}}$ was used to intensify the sampling of short events. By using the weighted average of the a priori variance $w_0$ from the $t_{\mathrm{E}}$ distribution of all events and the inverse variance from the PSPL-fit $w_{t_{\mathrm{E}}}=1/\sigma^2_{t_{\mathrm{E}}}$ we ensure that the sampling rate is not adjusted before the peak when the parameter uncertainty in $t_{\mathrm{E}}$ is high.
\begin{equation}
T \propto \sqrt{A} \frac{w_{t_{\mathrm{E}}} t_{\mathrm{E}}/\left<t_{\mathrm{E}}\right>+w_0}{w_{t_{\mathrm{E}}} +w_0} ,
\end{equation}

\section{Implemented strategy in 2013 and 2014}

\subsection{Observing constraints - pre-selection}

In order to achieve a sensible pre-selection of events, we use empirical thresholds based on the achieved signal-to-noise ratio during the start of the pilot phase:
\begin{equation}
t_{\mathrm{exp}} \approx \exp \left( 0.995  I_{\mathrm{mag}} - 11.042 \right) \,\mathrm{s},
\end{equation}
where the maximal exposure time was required to stay between 30 and 200\,s. The resulting expected accuracy has lead us to request events with a baseline $I_{\mathrm{mag}}<18$ and a predicted peak $I_{\mathrm{mag}}<17$. In addition, very long $t_{\mathrm{E}}>400\,\mathrm{d}$ events are excluded because they do not need to be observed at high cadence from follow-up observing teams. Most of the time we expect the telescope to be fully occupied with high priority targets and anomalies. If the event is well-past the peak, non-anomalous events with $u>1.5$ are dropped from further consideration. 

Independent of the constraints imposed by the system for automatic events, users can interact with the system through our online portal. As each site housed two or more telescopes, the target priority generator preferentially assigned events to telescopes that had already observed them. Additional observing resources on the 2m Faulkes telescopes could be activated by the person responsible for daily monitoring of operations. The exposure time and sampling rate for anomalous events was obtained from ARTEMiS \citep{dom08} and the SIGNALMEN anomaly detection algorithm, but exposure time estimates and sampling rates were adjusted for the 1\,m network. 

In order to assess how useful the suggested strategy is, we calculate the detection zone area as shown by \cite{hor09} and define the detection zone area as that area on the lens plane where a potential planet causes a deviation of at least 1.5\,\% of magnification $A$. That implicitly assumes that we are able to adjust our exposure times accordingly. For that purpose, we use typical event parameters from 2012 ($u_0=0.15, t_{\mathrm{E}}=25\,\mathrm{d}$) and introduce finite source effects as $\rho=10^{-3}$ Einstein radii for the binary model but not for the underlying point lens model. 

\subsection{Precedence for low-cadence fields}

Despite the increasing footprint of survey teams, there are patches on the sky that cannot be visited more than once or twice per night. Therefore, we observe events more or less intensively, i.e events that are visited at least 3 times per night from OGLE are observed only once a night from the 1\,m network to ensure early baseline coverage and reference frame. This rule only applies to regular events before reaching the peak and is our contribution to the microlensing community. 

\begin{figure}
\resizebox{\hsize}{!}{\includegraphics{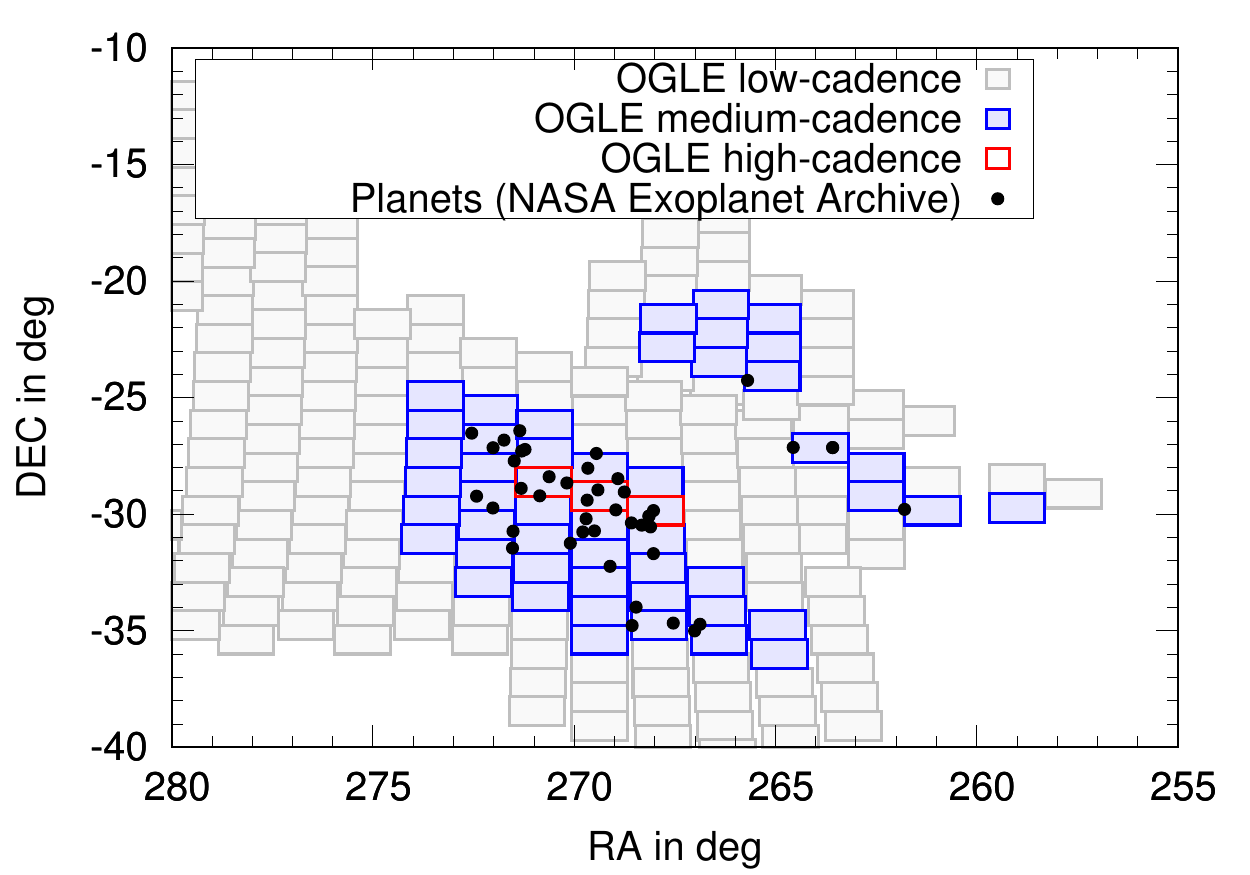}}
\caption{The number and location of planets confirmed by 2017 (NASA Exoplanet Archive) is shown along with the survey cadence.}
\label{fig:planetsevents}
\end{figure}

The priority flag requested by the network\footnote{see \url{http://robonet.lco.global/temp/tap1mlist.html}} is set to {\it low} for events with $\Omega_{s}<6$, to {\it medium} for $6 < \Omega_{\mathrm{s}} < 10$ and to {\it high} for $\Omega_{\mathrm{s}}>10$. The latter limit supersedes the nominal cadence by OGLE and ensures that all events can be observed on a regular basis. Events with $\Omega_{\mathrm{s}}<5$ are not queued at all. These limits do not refer to the current priority but to the maximum priority before uploading the target list. In the pilot phase of 2013, the estimates were made for a response time of 12 and later 6 hours. Since 2014, target lists are updated hourly. In addition to the 1\,m telescopes, RoboNet was awarded observing time on a robotic 2\,m telescope network consisting of the Faulkes Telescope North and South and the Liverpool Telescope. RoboTAP automatically requested 2\,m data in case of $\Omega_{\mathrm{s}}>30$ for 2013 and for a manual selection of the events in 2014.

The target list for each telescope or for the telescope network is populated after estimating the total available observing time on the network. Anomalous events are added until 30\,\% of the maximum available observing time is used up or one anomaly is requested per telescope. The latter can have a huge impact at the beginning and at the end of the season, where only a single anomalous event could be requested. The sampling interval is independent of anomalous events and is set on an event by event basis, partly automatically relying on the sampling rate recommended by ARTEMiS. At later stages team members can adjust the sampling interval depending on the events exact shape of the light curve and the expected duration of the event. There are automatic and public fitting systems such as RTModel\footnote{\url{http://www.fisica.unisa.it/GravitationAstrophysics/RTModel.htm}} helping to assess the nature of the event but at early stages the true nature remains uncertain.

The rest of the available time is populated with all other active events sorted by the aforementioned priority. In the pilot phase 2013 and at the beginning of the Key Project in 2014, there was not an immediate targets of opportunity (ToO) system available. For extremely magnified events or likely planets, continuous observations of a single target were requested on a single telescope, redirecting all other events to the remaining telescopes on a given site. Since 2014 we can request urgent observations as target of opportunity but due to the fast response of the scheduler (usually less than one hour) this is hardly ever necessary.

\section{RoboTAP in action}

RoboTAP generates target lists that are submitted to a target and observation manager code (ObsControl) which interfaces with the telescope network to request observations. In 2013 ObsControl read the telescope schedules directly per telescope and submitted microlensing observations into the gaps; once the network manager came online it ``just'' had to submit observation requests. The resulting requested observations were sent on a daily basis to the telescope network in 2013 and on an hourly basis in 2014. As soon as observations have been taken, the system informs the team about its current state. For the Key project the exact behavior of the scheduler changed slightly in the course of the project and no full-fleshed simulator of the scheduler was available. Instead, we have monitored how the scheduling worked in practice and if our desired target success rate was matched. 

\begin{figure}
\resizebox{\hsize}{!}{\includegraphics{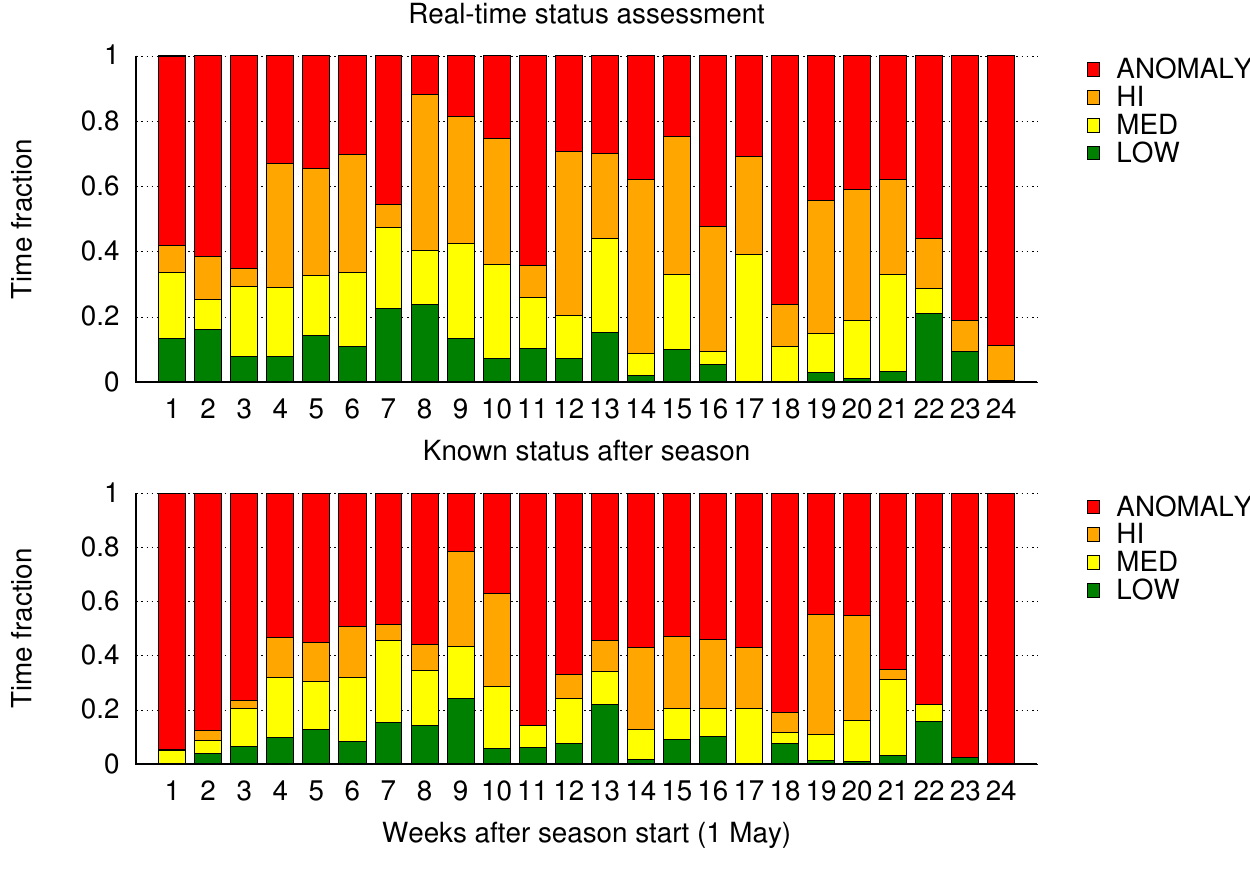}}
\caption{Weekly fraction of events of each category based on the real-time assessment of the system (top panel) and the actual priority levels based on the parameters and the event status obtained at the end of the season.\label{fig4}}
\end{figure}

The pilot phase season in 2013 lasted from 1st May until the end of September. During the season we logged the requested observing time by the target priority generator. Fig.~\ref{fig4} displays a weekly chart of how the real-time priority knowledge compares to the known final state of the events at the end of the season. As expected, some events that have been identified as regular events turn out to be anomalous, and particularly those of high priority. The actual fraction of anomalous events, including manually alerted ones differs a bit, because single anomalous events at the beginning and end of the season can override the fraction of anomaly-time we have envisaged. 
The agreement between the predicted priority level and the priority level for the final PSPL fit matches 82\% of the requested observing time for regular events. For anomalous events, a successful coverage with either anomalous or high-priority status was achieved in 89\% of the time of anomalous events. These high success rates testify to the efficiency of the algorithm.

The field of view of follow-up telescopes constrains the number of targets that can be followed-up. In the microlensing pilot program up to 12 targets can be observed at the same time and on average 7 events were on our target lists. Thresholds for rejecting events have been chosen in order to match the expected available observing time. The decision-making was relying on PSPL fits using the SIGNALMEN anomaly detector \citep{dom07}. For consistency, we have decided to focus our regular monitoring campaign on OGLE-IV\footnote{\url{http://ogle.astrouw.edu.pl/ogle4/ews/ews.html}} events. The network approach enables us to follow more than one strategy and events from other survey teams are separately followed as targets of opportunity.

\begin{figure}
\resizebox{\hsize}{!}{\includegraphics{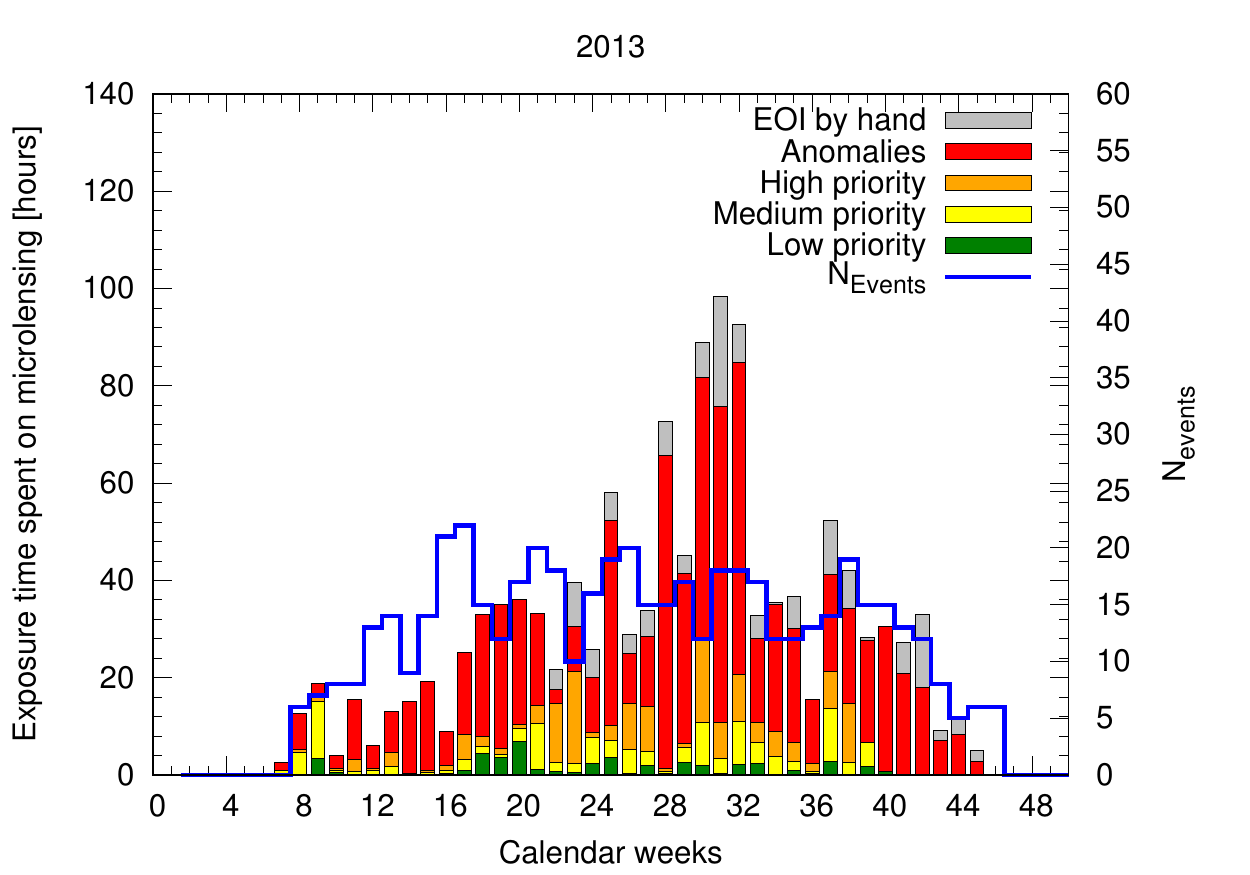}}
\resizebox{\hsize}{!}{\includegraphics{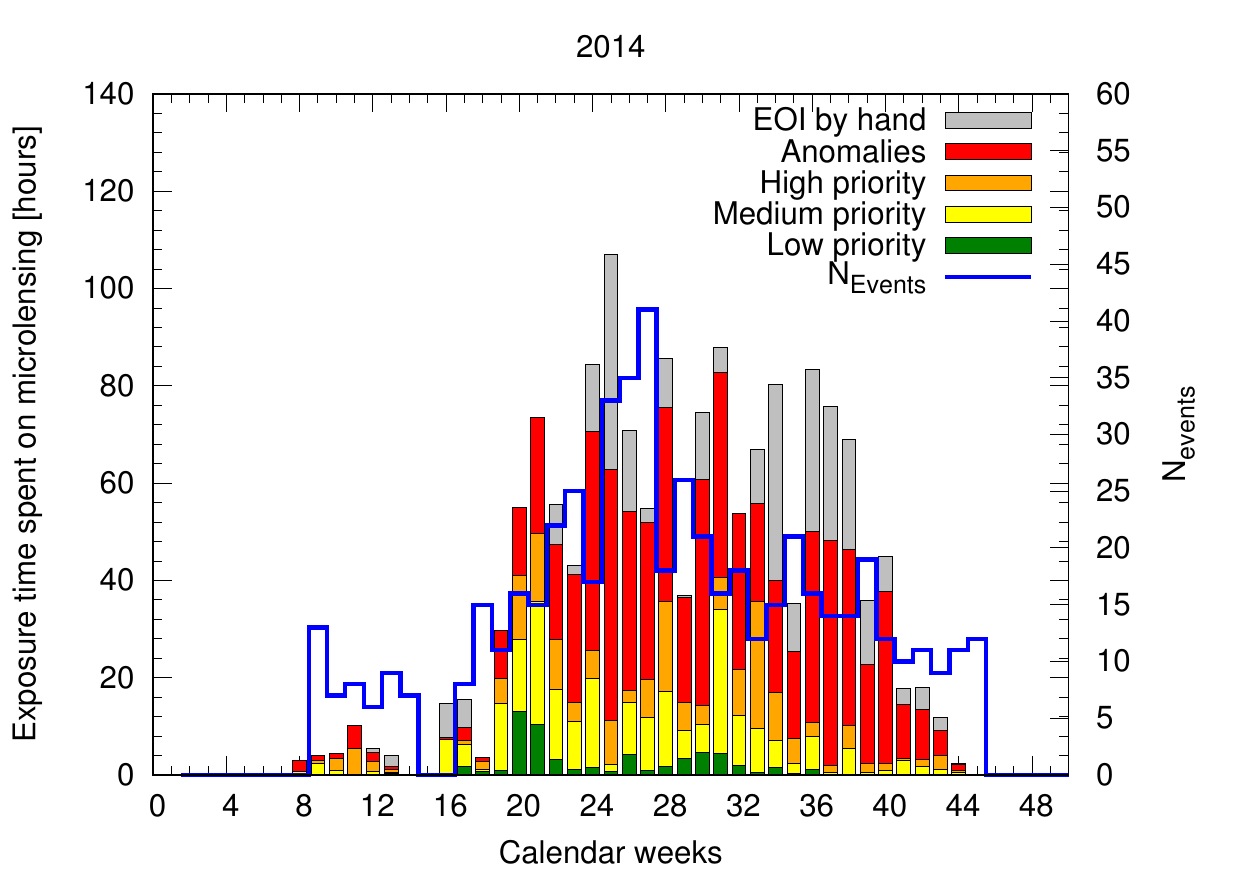}}
\caption{Weekly observations with the LCO 1\,m network of microlensing events in each category including manually requested observations, shortly denoted as events of interest (EOI), shown for the 2013 and the 2014 season. The continuous blue curve provides the number of followed events. The LCO scheduler came online since calendar week 18 in 2014.\label{fig:obs}}
\end{figure}

Fig.~\ref{fig:obs} shows the time per week spent on microlensing targets. The change in overall shape between the two seasons can be explained by the deployment of the network scheduler distributing observing time between different projects. Nevertheless, the same overall pattern is apparent in both years, with the largest fraction of observing time allocated to anomalous and high priority events.

\section{Expected number of planet detections}

To assess the number of planets that can be detected, we apply the following approach. Starting from the known PSPL model fit parameters for microlensing events in a given observing season, we employ the simplified Galactic model presented in \cite{tsa16} to derive fiducial physical masses and circular orbital separations. The sample of planets drawn is calibrated to the orbital and mass distribution given in \cite{cas12}. Our results are still consistent when simulating orbits $a \in \left[0.1,10\right]\,\mathrm{AU}$ and planet masses $M_{\mathrm{p}} \in \left[5,3178\right]\,M_{\oplus}$  distributed uniformly in $\log a, \log M_{\mathrm{p}}$.  

Depending on the expected host star mass, we randomly assume that a multiple-star is present. The multiplicity fraction of stars and brown dwarfs is roughly approximated by a linear decrease in logarithmic mass from 80\,\% for a $10\,M_{\odot}$ star to 20\,\% for a $0.1\,M_{\odot}$ which is loosely motivated by \citep{lad06,rag10}. For a uniform hourly sampling of events and based on the LCO exposure time calculator we obtain for the longest useful exposure time of 300\,s with the SBIG cameras 1.5\,\% for average conditions\footnote{\url{https://lco.global/files/etc/exposure_time_calculator.html}}. The actual uncertainty can be larger than that and we can still detect a signal but for uncertainties beyond 10\,\% effectively no planet can be detected any longer with the aforementioned criterion. As a side-remark, we are exclusively focusing on planets orbiting single-star hosts and exclude circumbinary planets \citep{ben16}. 

Instead of simulating complete binary-lens light curves, we inject planets into a point-source point-lens model and test those 7 points that have been observed closest to the anomaly to test if their cumulative $\Delta \chi^2>100$. In order to prevent single-outlier events, we request each observation to contribute by $\Delta \chi^2>100/7$. Requesting more points would exclude planets inducing only a short anomaly \citep{bea06} while discarding the threshold for each single observation would permit detecting events that are only weakly varying and difficult to distinguish from systematic or long-term effects such as annual parallax. 

If the expected annual sample of 2000 events could be used for our detections, we expect that 14 to 46 planets would be detectable orbiting single stars. The range of detectable planets corresponds to the $1\,\sigma$ uncertainty of \cite{cas12} and is used for all subsequent estimates. Only single-star hosts were considered and for that purpose 36\,\% of all planets were discarded because they are orbiting potential multiple stars and further 8\,\% were discarded because their host mass was below 70 $M_{\mathrm{Jupiter}}$. As a side-remark, the observed fraction of identified binary star lenses is 2\,\% \citep{tsa16} and that means that one third of our perceived single-lens events are actually multiple stars. Our most important idealization is an uninterrupted hourly sampling of all 2000 events with constant exposure time so that the SBIG cameras installed on the LCO 1m-network reach 1.5\,\% photometric accuracy for an $I=18\,\mathrm{mag}$ star. Our initial priority threshold for $\Omega_S$ reduces the number of detected planets by 15\,\% while reducing the requested time by a factor of 10. The invested time still exceeds the scope of a Key project by a factor of 50.

All requirements together exceed the capabilities of existing observing teams. Estimating the yield for the anomaly-triggered part is more difficult. Our former estimates give a hint on the fraction of constituents. Based on 10\% non-planetary anomalies and 2\,\% of anomalous events, we can expect to find a planet whenever 50 events have been sufficiently covered. 

\begin{figure}
\resizebox{\hsize}{!}{\includegraphics{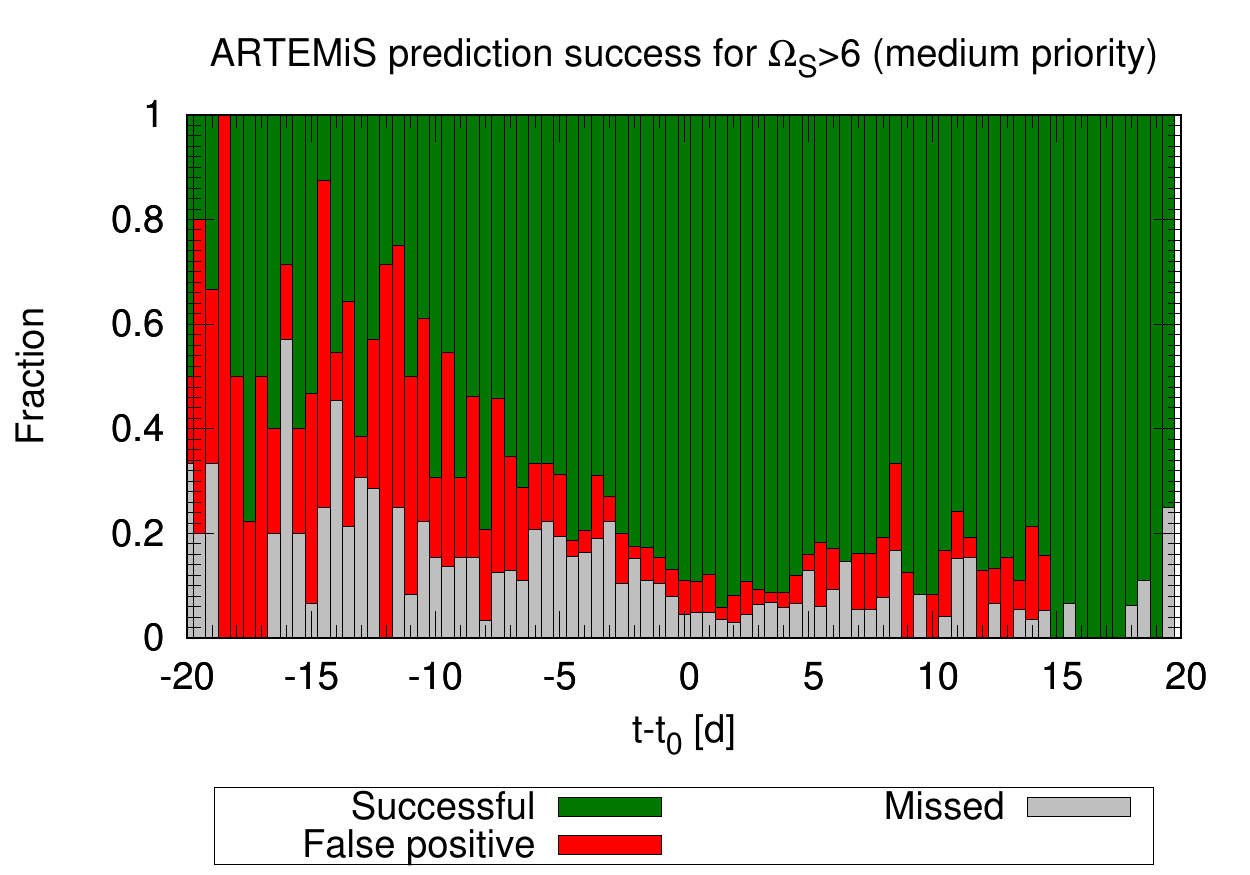}}
\caption{Successful predictions, false positives and missed predictions for PSPL parameters are shown as a stacked histogram, i.e. the color-shaded areas correspond to the relative fractions.\label{fig:evo}}
\end{figure}

We are still missing the contribution of the actual sampling strategy. For that purpose, we simulate light curves with the full range of sampling intervals and assign \,60\% of our allocated time for monitoring. So far we have assumed that the parameters of our light curves are appropriately prioritizing events at any given time. In practice, we are affected by uncertain event parameters. For a typical season, we determine the number of missed and incorrectly assigned priority categories as shown in Fig.~\ref{fig:evo}. The estimates allow on the event status at any given time compared to the final season parameter estimates. This leads us to a conservative estimate of 50\% loss for events selected before the peak and roughly 20\% after the peak. The expected planet yield decreases accordingly. In practice, the detection probability is higher in our selection of events and in the next section we will see how that supports our case.

\section{Achieved planet sensitivity}

From the observations taken in the seasons 2013 and 2014, we can estimate the expected number of planets from the non-anomalous part of our sample, using the end of season point-source point-lens parameter estimates. For the pilot phase 2013, we would have been sensitive to 2 to 8 planets (for the reported light curve uncertainties), while the sensitivity in the regular observing season 2014 was increased to 3 to 11 planets neglecting the aforementioned fraction of missed and overrated events. Both estimates assume the \cite{cas12} PMF and our 7-point detection criterion. 

We start by simulating planets uniformly in $\log q, \log s$ where the separation $s \in \left[0.1,10\right]$ Einstein radii and $q \in \left[0.03,10^{-5}\right]$ \citep{bon04}. A planet is detected whenever the resulting $\Delta \chi^2$ exceeds 100. Our reported photometric uncertainties are rescaled with one parameter so that $\chi^2_{\mathrm{red}}=1$ because our light curves can come from up to 8 telescopes. In addition, we keep the finite source radius constant for the binary model ($\rho=10^{-3}$) and compare it with a point-source point-lens model only. In a second step, we fit a point lens model to the simulated planetary light curve and check if the detection can be smoothed out and the $\Delta \chi^2$ can fall below 100. As a side remark, we have started the simulation using the ARTEMiS fit parameters which includes survey data. In the second step, we fit the model only to follow-up data. The detection is only accepted if it can be sustained by our follow-up data, but we do not claim that we can characterize the event without survey data. All accepted events are then divided by the number of samples per event and by that we obtain the expectation for the 118 regular events observed in the 2013-2014 season. Assuming each event contains one planet in the given range, we expect to find 4 in the season 2013 and 5 planets in the season 2014. Our earlier considerations are consistent as far as the expected planet yield is concerned. Defining an even more conservative threshold of $\Delta \chi^2>500$ as suggested by \cite{yee12} would reduce that by a factor of 1.4 and thus to 6 planets for both seasons combined.

\begin{figure}
\resizebox{\hsize}{!}{\includegraphics{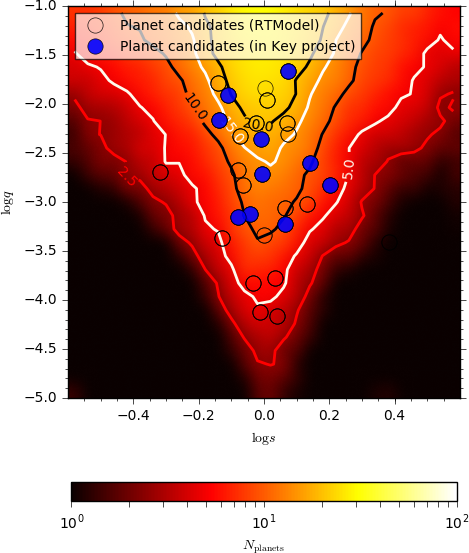}}
\caption{Key project sensitivity illustrated as expected number of detections assuming each event has a planet in the range $\log s \in \left[-1,1\right]$ Einstein radii and $\log q \in \left[\log 0.03,-5\right]$. To compare that with the number of detected planet candidates, we plot all planet candidates from real-time modeling efforts (RTModel) that have been reported 2013-2014 as part of the key project (filled) and all other planet candidates that were reported 2013-2016 (empty circles).\label{fig:sensplot}}
\end{figure}

In order to obtain an immediate impression how many planet candidates we have actually covered, we plot all events classified as planet candidates found on the RTModel web-page \footnote{\url{http://www.fisica.unisa.it/GravitationAstrophysics/RTModel.htm}} in Fig.~\ref{fig:sensplot}. The number of covered events exceeds one planet per year, but half of the events were observed only after receiving an anomaly trigger which also roughly matches the time allocation. The achieved planet detections nearly matches our expectation for the underlying regular event distribution and is consistent with \cite{cas12}.

In the season 2015, a non-negligible fraction of observing time was spent on following Spitzer targets. This gives us the opportunity to see how a season with a partly different sampling strategy behaves. In order to observe all Spitzer targets, the observing cadence was gradually reduced in order to secure points on all ongoing events. In principle, that should double our planet sensitivity, but in practice we see a similar number of covered planet candidates and one event with crucial data for characterizing the event \citep{str16}. The unusually good coverage on that event also enables us to reach our target mass ratio of $10^{-4}$.

\begin{figure}
\resizebox{\hsize}{!}{\includegraphics{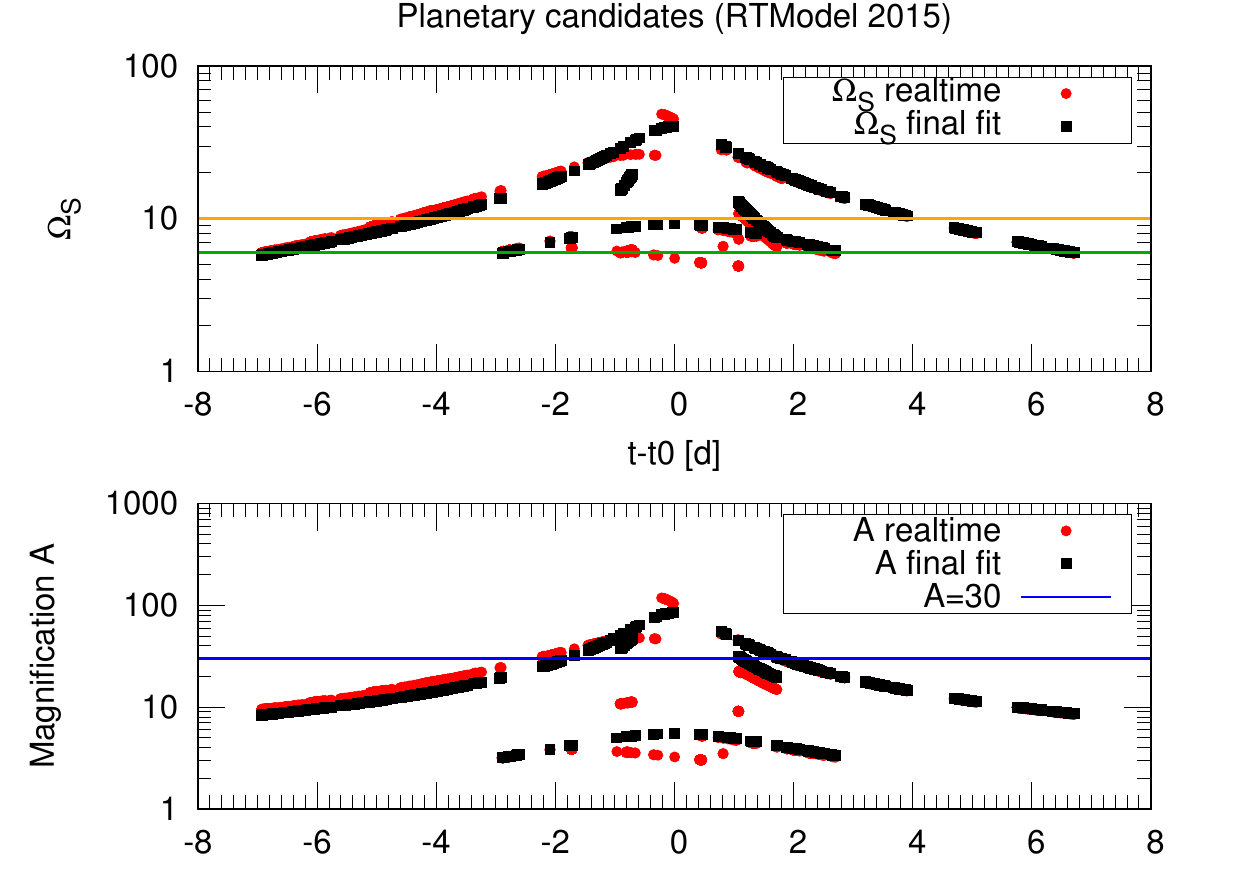}}
 \caption{The priority function and magnification is shown for those planet candidates that have been identified by RTModel (2015). High priority would correspond to a priority value above 10 and means that the event is observed.}
  \label{fig:planets2015}
\end{figure}

 Independent of the exact sampling strategy required, we show in Fig.~\ref{fig:planets2015} how the actual assessment of the priority has changed over time and what the actual priority should have been based on the final parameter estimates. All targets shown in Fig.~\ref{fig:planets2015} have been reported online by the real-time modeling platform RTModel\footnote{\url{http://www.fisica.unisa.it/GravitationAstrophysics/RTModel/2015/RTModel.htm}} and observed by RoboNet. The priority thresholds discussed and chosen in this work are evidently allocating more observing time to events with lower magnification ($<50$).

\section{Conclusions}

\begin{figure}
\resizebox{\hsize}{!}{\includegraphics{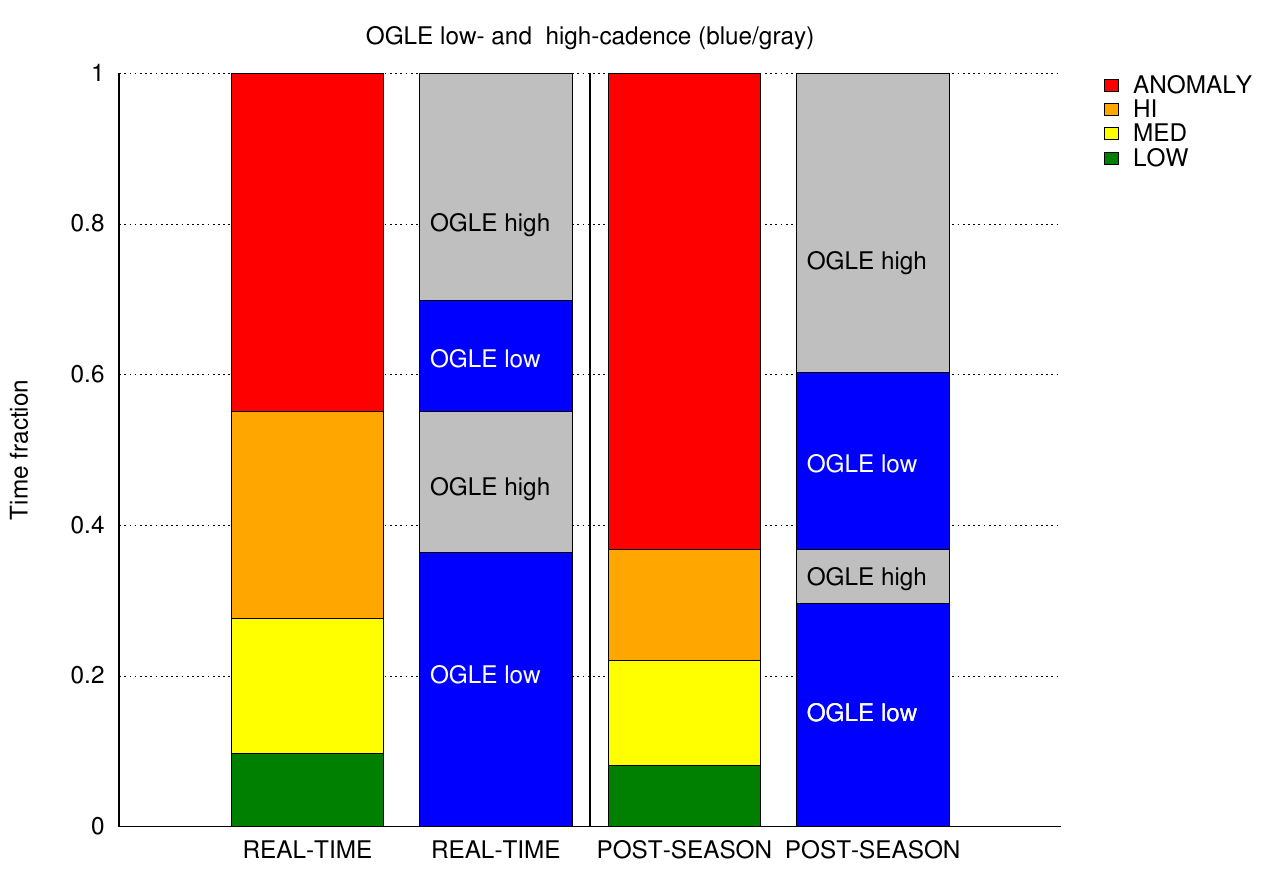}}
\caption{Time fraction of different priority levels for regular and anomalous events based on the observing logs (left) and recalculated for parameters at the end-of-season (right). The figure highlights how many anomalous events have been identified as regular, high priority targets. It also suggests that events from high-cadence survey fields are more likely to be assessed as an anomalous event.  \label{fig:category_match}}
\end{figure}

We have reported how follow-up observations of gravitational microlensing events can be automatically requested by our RoboTAP system. In that context, RoboTAP acts as an example of a broker between an incoming event stream and a robotic telescope network such as the LCO 1\,m network. The 2013 pilot phase provided some valuable and well-tested insights into the way targets can be selected. One of the insights shown in Fig.~\ref{fig:category_match} is our large contribution of observing time to help cover regular (PSPL-like) events in low-cadence fields, which are fields observed less than once per night by survey teams. Similar considerations will be required for all-sky surveys and their varying field cadence.

In the broader context of observing transient events triggered by future survey telescopes, like the ZTF and the LSST, we have shown that respecting the survey cadence is very useful as a selector: it helps to avoid regions where surveys achieve sufficient coverage for the respective science goal. At the same time targets should not be selected when there is only little chance to trigger an anomaly in due time. In retrospect, 80\% of our coverage went to regular targets. It is also noteworthy that regular events in the high-cadence zone in the real-time assessment more often turned out to be anomalous events and thus half of all observed events exhibited some form of anomaly.

Restricting follow-up observations to microlensing fields that survey teams can only visit once a night is not a viable option because event parameters and anomaly triggers remain elusive. For the fields defined in this work, the number of potential planetary events is reduced by a factor of 2. Two factors contribute to that: larger event uncertainties, and thus inappropriately low or high priority levels, and the differing blend ratio. That former is also an indicator for the number of available source stars.

\begin{figure}
\resizebox{\hsize}{!}{\includegraphics{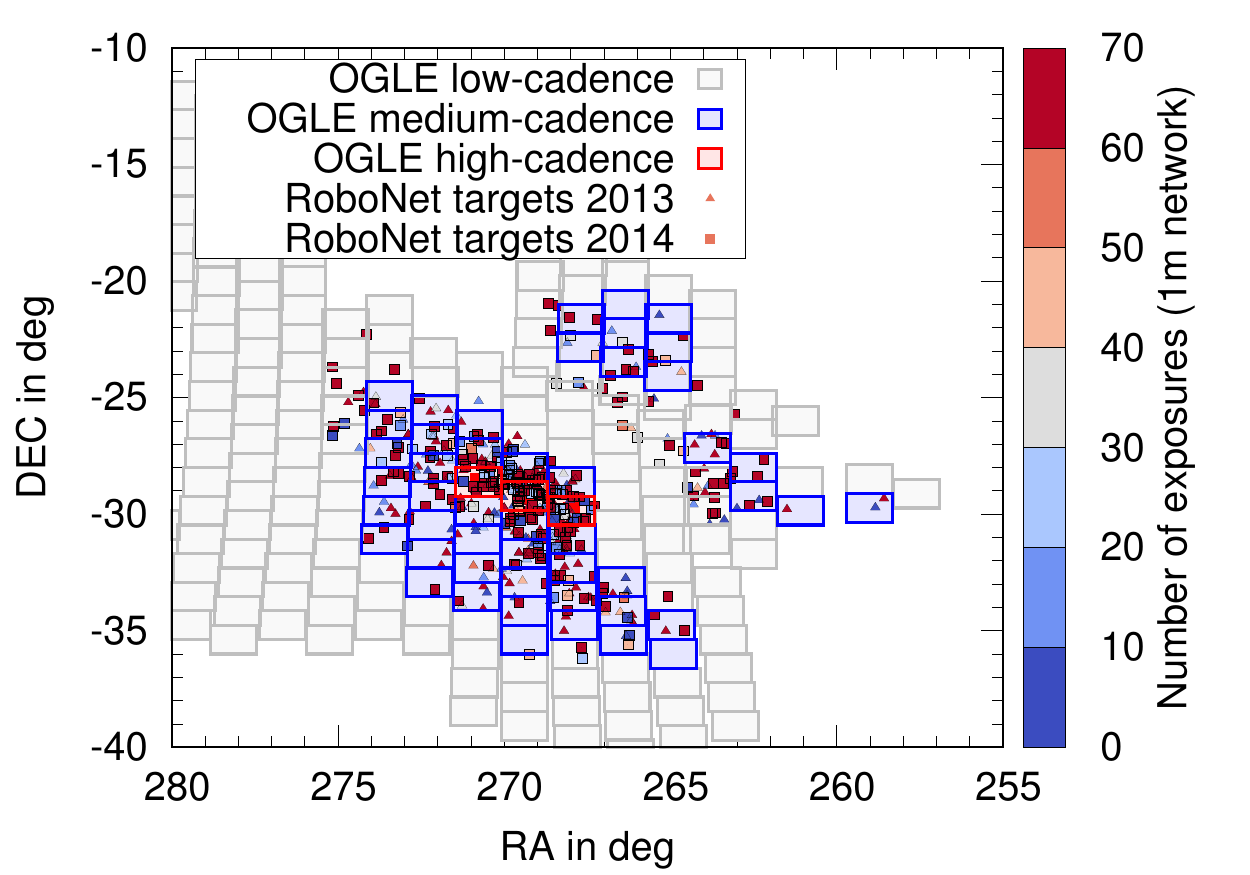}}
\caption{The distribution of observed events and their number of visits during the RoboNet campaign 2013-2014 is shown along with high-, medium- and low-coverage zones of the OGLE-IV survey. \label{fig:robo_coverage}}
\end{figure}

The anomaly detector requires multiple data points from two consecutive nights to deviate. Whenever surveys achieve only 1-3 visits per night, it is beneficial to follow anomalies from all-fields. The overhead of low-cadence follow-up for the initial phase of the Key project is shown in Fig.~\ref{fig:robo_coverage}. Starting in 2015, we have changed the interpretation of what belongs to each cadence category. Only 3 fields with the very highest coverage by OGLE are subject to a reduced cadence during regular follow-up because we have concluded that OGLE and other survey teams will be able to characterize these targets by themselves. 

Selecting microlensing events with high planet detection probability automatically ensures that the potentially limiting parameters for the planet mass ($q,t_{\mathrm{E}}$) are sufficiently characterized. This is another way of expressing that some change of brightness that permits us to find a planet also contributes to the characterization of its parameters. However, a better coverage in the wings is the preferable choice for the Einstein time. For future observing programs we suggest to study in more detail how to routinely include estimates of the expected achievable parallax signal in the selection process in a way that avoids a selection bias tending towards long duration events. Nevertheless, with current technology only half of the planets will benefit from that. In all other cases, we expect our approach to be appropriate.

When comparing the requested and observed fractions of all observing requests submitted to the telescope scheduler, we find that in 64\% of all cases the requested next observation was either immediately after the requested sampling interval or at most 15\,min later. That would be sufficient for detecting low-mass planets, but is not representative of present and future systems due to modifications in the target scheduling and a larger number of parallel projects. Our Key project sensitivity suggests that low-mass planets with $q<10^{-4}$ should have been detected, but actual candidates are above a mass-ratio of $q>10^{-3.5}$. Achieving a higher cadence, a higher photometric accuracy and covering more events at the same time is usually hard to accomplish. The Spitzer campaign in 2015 was scientifically fruitful, not least because of the planet discovery of OGLE-2015-BLG-0966 \citep{str16}. At the same time our nominal chance to detect planets was larger while the actual number of planet detections was reduced.

As far as finding and characterizing cool planets is concerned,
we expected to detect $\sim$5 planets per year. This estimate 
was consistent with the planet candidates identified in real-time 
modeling. Half of these candidates were identified in events with
anomaly-triggered observations, while the other half were found
in regularly monitored events. Further refinement of the results
reported here will come from analyzing the data obtained by 
the next microlensing Key Project on the LCO network (2017-2020).

\begin{acknowledgements}

This publication was made possible by NPRP grant \# X-019-1-006 from the Qatar National Research Fund (a member of Qatar Foundation).
Work by SM has been supported by the Strategic Priority Research Program ``The Emergence of Cosmological Structures" of the Chinese Academy of Sciences Grant No. XDB09000000, and by the National Natural Science Foundation of China (NSFC) under grant numbers 11333003 and 11390372. 
Cl\'ement Ranc's research was supported by an appointment to the NASA Postdoctoral Program at the NASA Goddard Space Flight Center, administered by Universities Space Research Association under contract with NASA. 
KH acknowledges support from STFC grant ST/M001296/1. 
This work makes use of observations from the LCO network, which includes three SUPAscopes owned by the University of St Andrews. The RoboNet programme is an LCO Key Project using time allocations from the University of St Andrews, LCO and the University of Heidelberg together with time on the Liverpool Telescope through the Science and Technology Facilities Council (STFC), UK.
This research has made use of the NASA Exoplanet Archive, which is operated by the California Institute of Technology, under contract with the National Aeronautics and Space Administration under the Exoplanet Exploration Program. 

We would like to thank the anonymous referee for his/her helpful comments and suggestions. Finally, we would like to thank all teams - and in particular the survey teams OGLE and MOA - for making their real-time data public.

\end{acknowledgements}


\end{document}